\documentclass[twocolumn,showpacs,preprintnumbers,prl,aps]{revtex4}

\newcommand{\BE}{\begin{equation}}
\newcommand{\EE}{\end{equation}}
\newcommand{\BA}{\begin{eqnarray}}
\newcommand{\EA}{\end{eqnarray}}
\newcommand{\BW}{\begin{widetext}}
\newcommand{\EW}{\end{widetext}}
\input epsf.tex

\begin{document}



\title{Frozen water waves over rough topographical bottoms}

\author{Liang-Shan Chen$^1$}\author{Zhen Ye$^{1,2}$\footnote{Corresponding author:
zhen@phy.ncu.edu.tw}} \affiliation{$^1$Department of Physics,
Fudan University, Shanghai, China and $^2$Wave Phenomena
Laboratory, Department of Physics, National Central University,
Chungli, Taiwan, China}

\begin{abstract}

The propagation of surface water waves over rough topographical
bottoms is investigated by the multiple scattering theory. It is
shown that the waves can be localized spatially through the
process of multiple scattering and wave interference, a peculiar
wave phenomenon which has been previously discussed for frozen
light in optical systems (S. John, Nature {\bf 390}, 661, (1997)).
We demonstrate that when frozen, the transmission of the waves
falls off exponentially, and a cooperative behavior appears, fully
supporting previous predictions. A phase diagram method is used to
illustrate this distinct phase states in the wave propagation.

\end{abstract}


\maketitle

Multiple scattering occurs when waves propagate in media with
scatterers, leading to many interesting phenomena such as the
bandgaps in periodically structured media and Anderson
localization in disordered media\cite{Anderson}. Within a bandgap,
waves are evanescent; when localized, they remain trapped
spatially until dissipated. The phenomenon of bandgaps and
localization has been both extensively and intensively studied for
electronic, electromagnetic, and acoustic systems\cite{Sheng}.

Recently, Wiersma {\it et al.}\cite{Wiersma} have demonstrated
that light can be forced to stand still in strongly scattering
semiconductor powders. The authors have shown a transmission
transition from classical diffusion to localization. When
localized, the optical transmission decays exponentially instead
of linearly with the thickness of a sample. At the transition, the
transmission has a power-law dependence on the inverse thickness.
This experiment not only leads to new applications\cite{John} in
optical data processing and laser action, but also sheds new
lights to the understanding of the Anderson localization
transition. Although regretfully not being shown by this
experiment, it has been further predicted by John\cite{John} that
a coherent behavior should appear for localized waves. In
addition, in the classical diffusion through a scattering medium,
the intensity distribution will fluctuate significantly due to
wave interference, while the fluctuation will be reduced in the
localized state.

Here we report that the Anderson localization phenomenon may also
be observed for the surface water wave propagation over random
bottoms, also termed as gravity waves in fluid mechanics. We show
that gravity waves can come to a complete halt in the presence of
random multiple scattering and wave interference. In the localized
state, not only the wave intensity decreases exponentially in
agreement with the observation in Ref.~\cite{Wiersma}, but also a
distinct phase coherence behavior prevails. The transition from
the diffusive to localization regime is signified by the
variations in the fluctuation of the transmission. The
localization regime overlaps partially with the bandgap of
corresponding regularly structured bottoms. These observations
fully support the predictions in Ref.~\cite{John}, indicating that
these phenomena are a generic property of waves.

The propagation of water waves over topographical bottoms has
actually been a subject of much research, from both ocean
engineering and fundamental research perspectives (e.~g.
Refs.~\cite{Papan,Chou,APL,Mc,Mei}). A comprehensive summary and
reference can be found in excellent textbooks
\cite{Lamb,CCM,Ding}.

The concept of Anderson localization has also been extended to the
study of the propagation of surface water waves over rough
bottoms. In 1983, Guazzelli {\it et al.} \cite{Guazzelli}
suggested that the phenomenon of Anderson localization could be
observed on one dimensional shallow water waves, when the bottom
has random structures. Later, Devillard {\it et al.} reconsidered
the problem by the potential theory \cite{Devillard}. The
experimental observation of water wave localization has been
subsequently suggested by Belzone {\it et al.}\cite{Belzons}.
These earlier attempts have been limited to the transmission
measurement and to one dimension.

A recent experiment\cite{Nature} has used water waves to
illustrate the Bloch wave phenomenon over a two dimensional
periodic bottom. This pioneering experiment has made it possible
that the abstract concept be presented in an unprecedentedly clear
manner.

The experimental advantures\cite{Belzons,Nature} pave a new avenue
for investigating the phenomena of Anderson localization in
disordered media and wave bandgaps in periodic structures. These
developments are the motivation for us to explore some important
properties of Anderson localization in the propagation of surface
water waves over random topographical bottoms.

\begin{figure}[hbt]
\epsfxsize=2.5in\epsffile{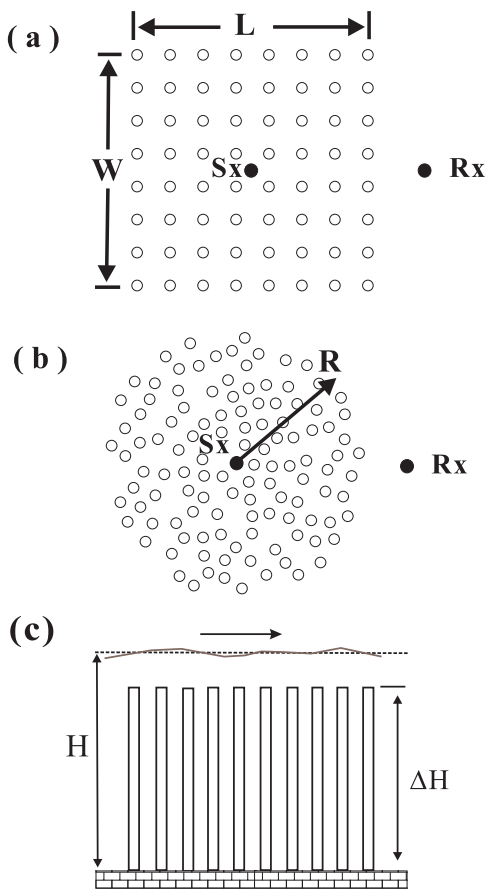} \caption{The conceptual layout
of the systems: (a) and (b) show the bird's views; and (c) the
side view. The circles denote the cylindrical steps, while Sx and
Rx refer to the stimulating source and the receiver
respectively.}\label{fig1}
\end{figure}

Making the results experimentally testable, we will adopt the
systems from the experiment\cite{Nature}. The conceptual layout of
the systems is illustrated in Fig.~\ref{fig1}. We consider a water
column with a uniform depth $H$. There are $N$ cylindrical steps
mounted on the water bottom. For simplicity, the steps are assumed
to be identical. The heights of the steps are denoted by $\Delta
H$ and the radii are $a$; we can also extend to consider the
situation of cylindrical dimples by letting $\Delta H <0$. For
comparison, we will consider both the randomly and the
corresponding orderly arrangements of the steps on the bottoms. In
the ordered case, the steps form a square lattice with the lattice
constant $d$; therefore the areal occupation fraction by the steps
is $\beta = \frac{\pi a^2}{d^2}.$ In the random case, the steps
are placed completely randomly within a circular area of radius
$R$. In both cases, the areal occupation fraction is same;
therefore $L = \sqrt{(N\pi a^2)/\beta}$ and $R = \sqrt{(N
a^2)/\beta}$. A monochromatic transmitting source of angular
frequency $\omega$ is located in the middle of the arrays of the
steps. The transmission is measured by a receiver located outside
the arrays. The water surface is in the $x-y$ plane. In the
simulation, all lengths are scaled by the lattice constant $d$.
This is a two dimensional problem.

The governing equations for the motion of surface water waves in
the systems described by Fig.~\ref{fig1} can be obtained by
invoking the Newton's second law and the conservation law of mass.
The formulation has been given in Ref.~\cite{Nature,PRE} and the
detailed derivation is given in Ref.~\cite{Ye}. Here we only list
the final equations.

The displacement of the water surface is denoted by $\eta(\vec{r},
t)$. The equation of motion for its Fourier component is derived
as \cite{Nature,PRE,Ye} \BE
\nabla\left(\frac{1}{k^2}\nabla\eta(\vec{r})\right) +
\eta(\vec{r}) = -4\pi\delta(\vec{r}-\vec{r}_s),
\label{eq:finalb}\EE where $\nabla =
\partial_x \vec{e}_x +
\partial_y \vec{e}_y$, $\vec{r}_s$ is the location of the transmitting source, and the wavenumber $k$ satisfies \BE
\omega^2 = (gk(\vec{r}) +
bk^2(\vec{k}))\tanh(k(\vec{r})h(\vec{r})),\label{eq:dispersion}\EE
in which $b$ is the capillary length. For a fixed frequency
$\omega$, the wavenumber varies as a function of the depth
$h(\vec{r})$. Without the steps, the wave field is $\eta_0$. We
note that Eq.~(\ref{eq:finalb}) is derived when non-linear effects
are ignored. The non-linearity may give rise to
delocalization\cite{Mei}.

In this paper we will apply Eq.~(\ref{eq:finalb}) to the systems
depicted in Fig.~\ref{fig1}. The transmitted waves will be
scattered repeatedly by the steps, forming an orchestral pattern
of multiple scattering. Such a multiple scattering process can be
solved for any arrangement of the steps by the multiple scattering
theory\cite{Ye} following the work of Twersky\cite{Twersky}. In
the computation, the transmission is normalized such that it is
unity when there are no scatterers, thus eliminates the trivial
geometrical spreading effect In the periodic case, the plane wave
expansion method will be used to compute the band structures of
the water waves\cite{Ye}.

A set of numerical simulations has been carried out. In the
simulation, the following parameters are adopted from the
experiment\cite{Nature}: the depth $H = 2.5$ mm; the height of the
cylinders $\Delta H = 2.49$ mm; in the periodic case, the lattice
constant $d = 2.5$ mm; the radius of the steps $a = 0.75$ mm; the
capillary length $b = 0.93$ mm. In the random case, the
transmission intensity is averaged over the random configurations.

\begin{figure}[hbt]
\epsfxsize=2.5in\epsffile{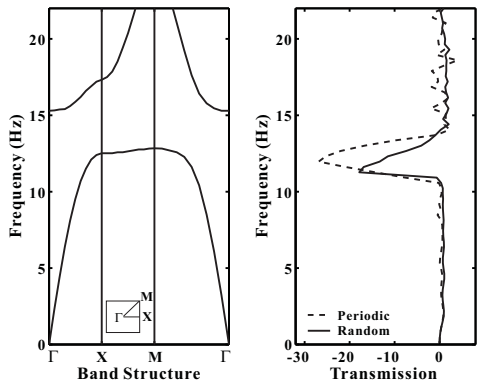} \caption{Right Panel:
Normalized transmission $\ln|T|^2$ versus frequency for the
complete random arrays (solid line) and the corresponding square
lattice of cylindrical steps (dotted line). The transmission in
the periodic case is along the $\Gamma X$ direction. Left panel:
the band structures for the square lattice. The inserted box in
(a) denotes the Brillouin zone.}\label{fig2}
\end{figure}

First, in Fig.~\ref{fig2} we show the normalized intensity of
transmitted waves ($|T|^2 = |\eta/\eta_0|^2$) as a function of
frequency, and the band structure of the corresponding square
arrays of the cylindrical steps. When computing the transmission,
the receiver is located at two lattice constants away from the
arrays. To ensure the stability of the results, enough modes and
number of steps have been considered. For instance, the maximum
mode number and the maximum array size considered are 9 and
14$\times$14 respectively. The wave transmission along the $\Gamma
X$ direction is shown for the periodic case. The transmission
through the random arrays of steps is also plotted. In both cases,
the source is placed in the middle of the arrays.

Here we see that a complete bandgap region can be identified as
ranging from about 12.5 to 16 Hz. A strong localization regime is
shown to range from 10 to 14 Hz in the random case. Though
shallower, this regime overlaps with the inhibition regime in the
periodic case. By comparison, we see that although not exactly
overlapping with each other, the strong localization and complete
bandgap regime are close to each other; thus finding the complete
bandgaps will facilitate locating the localization regimes, as
suggested in Ref.~\cite{JohnPRL}. We note here that in the
periodic case the inhibition regime in the transmission does not
fully overlap with the complete bandgap. This is because we have
put the source inside the array. When the transmission is measured
across the sample, i.~e. the source is placed on one side of the
sample and the receiver is placed on the other, the two will
overlap.

To show the conjectured frozen feature of localized waves and
fluctuating feature of non-localized waves\cite{John}, we have
considered two frequencies: one is in the passing band 5 Hz, and
the other is with the complete bandgap region, 11.966 Hz. We write
the wave field $\eta(\vec{r})$ as
$|\eta(\vec{r})|e^{i\theta(\vec{r})}.$ This allows us to separate
the amplitude field and the phase field of the waves. For the
phase field $\theta$, we define a unit phase vector field as
$\vec{u} = \cos\theta \hat{e}_x + \sin\theta \hat{e}_y.$ Both
phase vector field and the wave intensity field $|\eta|^2$ can be
plotted in the $x-y$ plane. The significance of the phase vector
field is as follows. The intensity flux of surface water waves can
be shown as $\vec{j} \sim |\eta|^2\nabla\theta$. It is clear that
when $\theta$ is constant, at least by spatial domains, while
$|\eta| \neq 0$, the wave transport would come to a stop and the
waves will be localized or frozen in the space.

\begin{figure}[hbt]
\epsfxsize=3.35in\epsffile{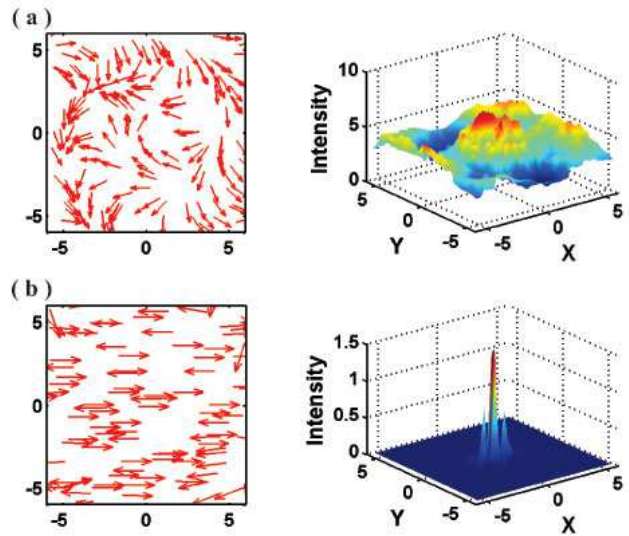} \caption{The phase diagram
and spatial distribution of the intensity field $|\eta/\eta_0|^2$
at two frequencies for one random configuration: (a1) and (a2) $f$
= 5.0 Hz; (b1) and (b2) $f$ = 11.966 Hz. Left panel: the phase
diagram for the phase vectors defined in the text; here the phase
of the source is set to zero. Right panel: the intensity spatial
distribution in the horizontal plane.}\label{fig3} \end{figure}

In Fig.~\ref{fig3}, the two-dimensional spatial distribution of
the normalized intensity $|\eta/\eta_0|^2$ and the phase vector
field are plotted for the two frequencies. The phase vectors are
located randomly in the $x-y$ plane. For $f$ = 5 Hz, the intensity
spreads spatially, meanwhile the phase vectors point to various
directions. This indicates that waves are not yet localized at
this frequency. The results in Fig.~\ref{fig3}(b) nicely
demonstrate the properties of the localized or frozen waves. At
$f$ = 11.966 Hz, the wave intensity is mainly confined near the
transmission site. Meanwhile, there is an ordering in the phase
vector field, that is, all the phase vectors either point to the
same direction or the opposite direction, indicating that the
phase field is constant by domains. Here we clearly demonstrate
that the localized waves behave as a standing wave in the random
media. These observations fully comply with the above general
discussion of localization, and also support the previous
predictions. The intensity distribution in Fig.~\ref{fig3}
realizes remarkably well what has been conjectured in Fig.~1 of
Ref.~\cite{John}. We note that the disorientation at the boundary
is due to the finite size in the simulation. For a finite system,
the wave can leak out at the boundary, resulting in disappearance
of the phase coherence. When enlarging the sample size, we observe
that the area showing the perfect phase coherence will increase.
We have further verified that the features shown in
Fig.~\ref{fig3} remain quantitatively the same for any other
random configuration.

\begin{figure}[hbt]
\epsfxsize=2in\epsffile{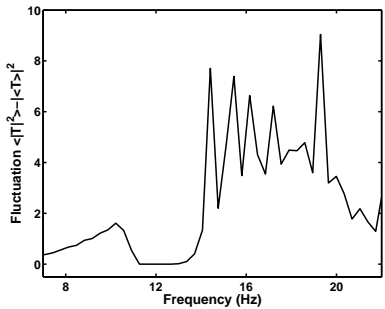} \caption{The fluctuation of
transmission as a function of frequency. }\label{fig4}
\end{figure}

We have also considered the fluctuation in the transmission.
Fig.~\ref{fig4} plots the fluctuation versus frequency for the
random case. The sample size $R$ is about 9 d. Here we see that
the fluctuation tends to be zero within a regime which is
consistent with the strong localization range discussed for
Fig.~\ref{fig2}. At around the localization transition edges,
significant peaks in the transmission fluctuation appear. For
extremely low frequencies, the fluctuation tends to disappear.
This is because that at when the frequency approaches zero, the
scattering strength will diminish, and thus the wave propagation
will be no longer affected by the steps.

\begin{figure}[hbt]
\epsfxsize=3.25in\epsffile{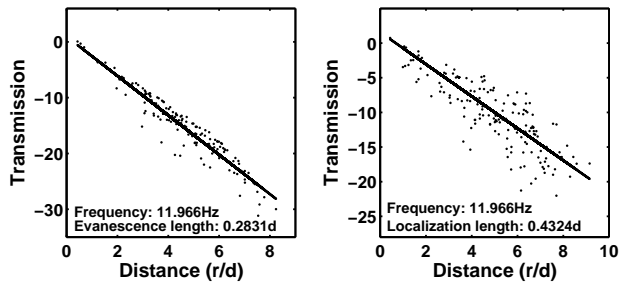} \caption{Wave transmission
versus the distance away from the source at 11.966 Hz: (a) the
periodic case; (b) the random case. The geometrical spreading
factor has been removed by the normalization. The slopes are
fitted from the least square method.}\label{fig5}
\end{figure}

To find the localization length, we plot the wave transmission in
all directions as a function of the distance from the source. The
results are shown in Fig.~\ref{fig5}. For comparison, the periodic
case is also potted. Here, the simulation data are shown by the
black squares, and the results fitted by the least square method
is shown by the solid lines; the deviations from the lines reflect
the inhomogeneity. It is shown that after the removing the trivial
geometrical spreading factor, the data can be fitted well by the
exponential function $e^{-r/\xi}$. From the slopes of the solid
lines, we obtain the evanescence length in the ordered case and
the localization length in the random situation respectively.

In summary, we have investigated the Anderson localization
phenomenon in the framework of gravity waves over rough bottoms.
As a comparison, the case of corresponding regularly structured
bottoms is also considered. The results indicate that when
localization occurs, the wave intensity is confined near the
transmitting point, and falls off exponentially. We have also
demonstrated that the localized waves stand still in the space,
represented by the phase coherence behavior. In addition, the
transition from classical diffusion to localization is associated
with a significant change in the transmission fluctuation. The
observation supports the previous predictions on localized waves.
Since water waves are a simple macroscopic system, experiments
could be readily performed. Therefore, many significant phenomena,
previously expected at microscopic scales such as the discussed
Anderson localization, may be demonstrated with water waves.

LSC is supported by the graduate program at Fudan University. The
encouragement and support from Prof. X. Sun are greatly
appreciated. ZY is grateful to the city of Shanghai for the Bai Yu
Lan Fund for visiting scholars.

\end{document}